\documentclass[aps,prl, reprint, twocolumn,  superscriptaddress]{revtex4}
\usepackage{graphicx}

\begin{document}
\title{Switching and amplification in disordered lasing resonators}

%
%
%
%
%
%

\author{Marco Leonetti}
\affiliation{ISC-CNR, UOS Sapienza, P. A. Moro 2, 00185 - Roma, Italy.}
\affiliation{Instituto de Ciencia de Materiales de Madrid (CSIC) Calle Sor Juana Inés de la Cruz 3, Cantoblanco 28049 Madrid Espa\~{n}a.}
\email[Corresponding Author: ]{marco.leonetti@roma1.infn.it}

\author{Claudio Conti}
\affiliation{Dep. Physics University Sapienza, P.le Aldo Moro 5, I-00185, Roma Italy}
\affiliation{ISC-CNR, UOS Sapienza, P. A. Moro 2, 00185 - Roma, Italy.}

\author{Cefe Lopez}
\affiliation{Instituto de Ciencia de Materiales de Madrid (CSIC) Calle Sor Juana Inés de la Cruz 3, Cantoblanco 28049 Madrid Espa\~{n}a.}

\begin{abstract}
\textbf{Controlling the flow of energy in a random medium is a research frontier \cite{Ohiorhenuan2010,Perseguers2010,jaeger96} with a wide range of applications \cite{Redding2012}. As recently demonstrated \cite{M.2010}, the effect of disorder on the transmission of optical beams\cite{ShengBook}, may be partially compensated by wavefront shaping\cite{Katz2012}, but losing control over individual light paths. Here we report on a novel physical effect whereby energy is spatially and spectrally transferred inside a disordered active medium by the coupling between individual lasing modes. We show that is possible to transmit an optical resonance to a remote point by employing specific control over optical excitations. The observed nonlinear transport bears some analogies to a field-effect transistor for light, which acts as a switch and as an amplifier.}
\end{abstract}

\date{\today}
\maketitle

Nowadays information is  processed either electronically\cite{Sze} or optically\cite{John:99,largescalePIC2011} by using devices like transistors and optical modulators, which are also developing in many unconventional fashions\cite{Jansen2012, Maeda2012,Hwang2009,LeMieux04072008,Fan27012012,O'Brien2003, Matsuo2010}. However the mechanisms underlying signal processing and transport in the presence of disorder are in many respects unknown even if they occur in numerous fields like biology, quantum networks, granular systems, and many others\cite{Ohiorhenuan2010,Perseguers2010,jaeger96}. There are several open questions regarding information transmission in complex systems, for example, which are the relevant degrees of freedom, which is the value of their mutual coupling and which is the way to route and control energy flow in presence of randomness.  In optics, random lasers (RLs) i.e., lasers based on disordered materials\cite{Wiersma_Rew, Cao03r, Leonetti2011}, are highly nonlinear, their modes exhibit variable spatial extent\cite{leonetti:051104}, and pronounced mutual interactions, thus they are a fertile ground where long range transport can be studied. In such open systems the various forms of localization\cite{Fallert_coexistence_nature} are naturally coupled thus it must be possible transferring a light from one mode to another.
Here we: I) identify coupled resonances located at distant locations in a RL; II) measure their effective coupling in an absolute way and III) demonstrate switching, routing and amplification of a signal from one mode to another. Moreover the results are theoretically interpreted in the framework of the coupled mode theory. The observed effect can be described in terms of an analogy between a field effect transistor and a RL with two pump wedges working as controls, showing that signal amplification and all-optical gating can be achieved.


\begin{figure}
\includegraphics[width=\columnwidth]{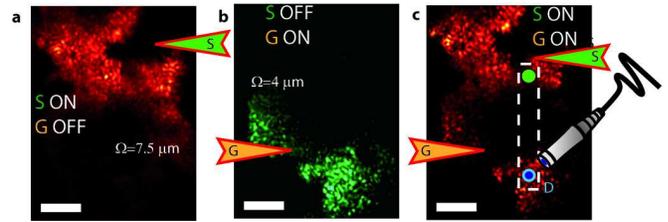}
\caption{\textbf{All-optical disordered-based transistor.} Optical image of the intensity emitted by the titanium dioxide cluster in various pumping configurations: \textbf{a} Source only, \textbf{b} Gate only and \textbf{c} Source and Gate simultaneously. In \textbf{b} the scheme of an electronic field-effect transistor is shown. Bar corresponds to 5 $\mu$m.
\label{fig:figure1}}
\end{figure}

The system considered is a cluster of titanium dioxide nanoparticles embedded in a rhodamine solution, which acts as the optically amplifying medium under excitation by a laser pump beam at wavelength $\lambda$ = 532 nm. This disordered device comprises many nanoparticles that self-assemble to form an optical resonator. By tailoring the spatial profile of the pump beam we control the activated modes \cite{leonetti2013active}: the directional pumping determines the spatial regions of the cluster which are pushed above threshold\cite{Leonetti_pra2011}, and their degree of localization, i.e., the spatial extension of the involved electromagnetic resonances or modes\cite{leonetti:051104}.

In our experimental setup (details in references \cite{Leonetti_pra2011, leonetti:051104}) the pump is spatially modulated so that selected local lasing resonances are activated. This is achieved by exploiting a scheme in which the shape of the pumped region generates stimulated emission engineered to hit the cluster at pre-defined points in pre-defined directions. The pumped area comprises two wedges of fixed orientation and controllable relative position and angular aperture(triangular arrows in Figure \ref{fig:figure1}) and a superimposed circular spot (not shown).

The two wedges are parallel, pointing in opposite directions, and vertically shifted by an amount $\Delta X$ ( $\Delta X$ = 15 $\mu$m in Figure \ref{fig:figure1}). In this way we are able to activate modes at a desired location and, for each mode, we control the localization length $\Omega$ by  the angular spread of the wedges\cite{leonetti:051104}.

To illustrate our results we employ the terminology of field-effect electronic transistors, where the passage of electrons from a region of the device where carriers are injected ({\it Source}), to a spatially separate region where carriers are collected ({\it Drain}), is controlled by the voltage applied to a third region ({\it Gate})\cite{Sze}. In analogy, we define three spatially separate regions of our cluster (named C1): the upper part is the \emph{Source} (S) area , the middle-lower part is the \emph{Gate} (G)area and the lower part is the \emph{Drain} area. \emph{Source}  and \emph{Gate} (figure \ref{fig:figure1}b) may be activated (ON state) by pumping the corresponding wedge shaped area, either individually (Figure \ref{fig:figure1}a for S and Figure \ref{fig:figure1}b for G) or simultaneously (Figure \ref{fig:figure1}c),  while \emph{drain} is used only as a probe, to collect light (with a fiber). Thus the cluster may be prepared in three different configurations: 1 \emph{(S:ON; G;OFF}) 2 \emph{(S:OFF;G:ON} and 3) \emph{(S:ON;G:ON)}. In addition light may be collected at the \emph{Source} (upper green spot in figure \ref{fig:figure1}c) or at the \emph{Drain} (lower blue spot in figure \ref{fig:figure1}c).


\begin{figure}
\includegraphics[width=\columnwidth]{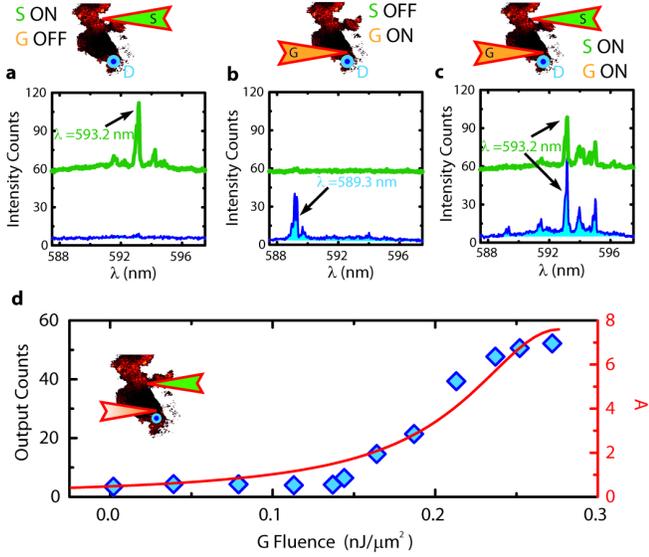}
\caption{ \textbf{Polychromatic action and amplification.} Spectra measured at S (green line) and at D (azure line) in different pumping configurations: \textbf{a} Source ON, Gate OFF; \textbf{b} Source OFF, Gate ON; \textbf{c} Source ON and Gate ON. Schemes atop each panel show the adopted configurations. \textbf{d} The intensity of the the S mode ($\lambda = 593.2$ nm) measured at D as a function of pump fluence at the Gate. The right y axis report amplification of the source signal measured as a function of the gate fluence while the red solid line is the best fit to a Lorentzian accounting for the mode coupling (see SI).
\label{fig:figure2}}
\end{figure}

The D and S signal from the cluster C1 pumped in three configurations are reported in Figure \ref{fig:figure2}. Configuration (S:ON;G:OFF) (panel \ref{fig:figure1}a) reports complectly different signals for S and D (in particular no signal is observed at D).


The opposite happens in Figure \ref{fig:figure1}b for configuration (S:OFF;G:ON): we find signal at the drain D but no signal at the source S. Furthermore, the signal at the Drain has a spectral content which bears no resemblance whatsoever with either spectrum in the previous configuration.

Finally in Figure \ref{fig:figure1}c, the S and G wedges are both pumped (S:ON;G:ON), and we observe that the signal at the Drain D has the same spectral content as that at Source S (Fig. \ref{fig:figure2}c) and no resemblance with its former lineshapes in 2a or 2b, providing evidence that light in S has been {\it routed} to the Drain region, with G acting as a gate.

Recapitulating we see that in the (S:ON;G:OFF) case, the S mode is clearly visible (showing a principal resonance at $\lambda = 593.2$ nm) with no emission at the D position while in the (S:OFF;G:ON) case a resonance at $\lambda = 589.3$ nm is found at the D position, with no emission from S. Hence, we find two modes located at different positions and with different spectra (with mode S being more spatially extended than mode D: localization length $\Omega$ is reported in figure \ref{fig:figure1}).

When both modes are activated at the same time (S:ON;G:ON) a striking effect is provoked: while S is oscillating at its own frequency (thus independent of the state of G), the D mode is forced to oscillate at the very same frequency of S ($\lambda = 593.2$ nm). The resonance lying at S retains its frequency but that at D can be driven by G to adopt the S spectral content. In other words, {\it the spectral content of S is transported to D, when the G is turned on}.

\begin{figure}
\includegraphics[width=\columnwidth]{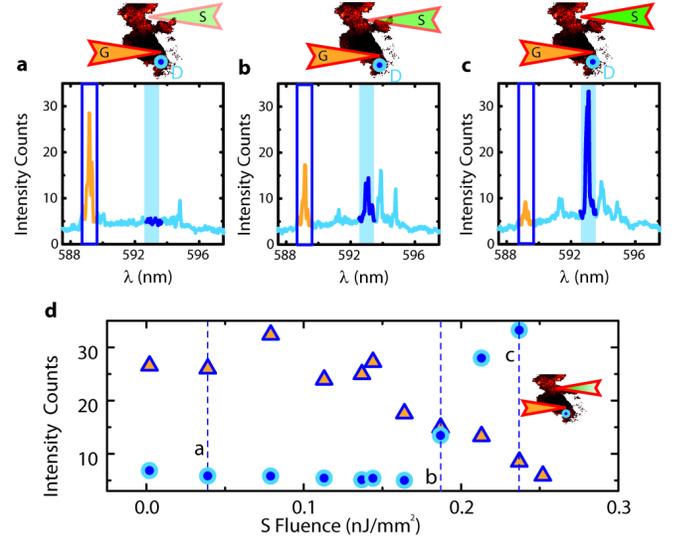}
\caption{\textbf{The switching process.} Spectrum obtained from the drain as a function of the energy at the source. The fluence at the source wedge is  0.04 nJ/$\mu$ m$^2$ in panel \textbf{a}; 0.18 nJ/$\mu$ m$^2$ in \textbf{b}, and 0.23 nJ/$\mu$ m$^2$ in \textbf{c}. Panel \textbf{d} shows the intensity of the D mode together with the G mode [the mode obtained for the (G ON, S OFF) condition] as a function of the fluence at S. Data corresponding to the top panels are indicated.
\label{fig:figure3}}
\end{figure}

This effect can be likened to an all-optical transistor, in which the energy flux can be routed. This is confirmed by the input/output curve shown in figure \ref{fig:figure2}d, which allows to extract an amplification (fig \ref{fig:figure2}d right axis), that grows with the control signal, as in electronic field effect transistor where the small signal amplification increases with the bias voltage at the gate terminal. We also stress that the process is inherently parallel, notice that several wavelengths are simultaneously routed from $S$ to $D$, as shown in the spectra in Fig.\ref{fig:figure2}c.

To understand the physical mechanisms underlying this controlled transport, we measured the spectrum at D as a function of the energy provided through the S wedge (when G is kept ON). Figures \ref{fig:figure3}a, b, and c show the retrieved average spectra (averaging over 25 single shots), revealing that the mode at D (excitable with the gate G:ON at $\lambda=589.3$nm, blue rectangle) is steadily suppressed as the S mode (at $\lambda=593.2$nm, azure shaded area) gains intensity. Figure \ref{fig:figure3}d, displays the average G (S) mode intensity as full triangles (open circles) as a function of the fluence supplied through the S wedge. Vertical lines labeled a, b and c correspond to spectra in figures \ref{fig:figure3}a, b and c. The progressive transfer of intensity from one mode to the other shows that, above a certain threshold, energy is routed from the G to the S mode due to the onset of the the nonlinear interaction, and the competition between the involved modes.

\begin{figure}
\includegraphics[width=\columnwidth]{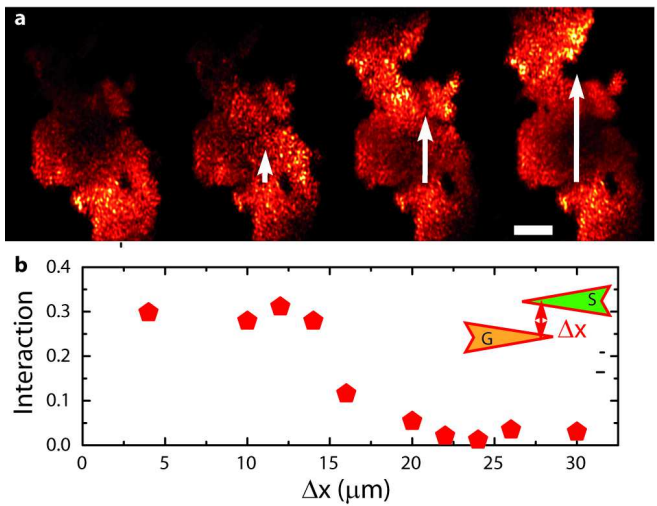}
\caption{\textbf{Range of the interaction.} \textbf{a} images of the emission from the cluster C1 for different values of $\Delta X$; \textbf{b} Interaction obtained from the spectral overlaps (see text) as a function of $\Delta X$. Bar corresponds to 5$\mu$m.
\label{fig:figure4}}
\end{figure}

The phenomenon may be explained exploiting the coupled mode theory equations \cite{Haus_Mode_Loking} for two individual interacting modes: a larger one (mode S, i.e. the source mode) and a smaller one (mode D, i.e. the drain mode). At small pumping for the Drain mode, the equation for its complex amplitude reads:
\begin{equation}
\frac{d a_D}{d t}=-i\omega_D a_D +[g_D-\alpha_D]a_D +K_{SD}a_S,
\end{equation}
where $a_D$, $\omega_D$, $g_D$, $\alpha_D$, $K_{SD}$ are respectively the amplitude of the mode D, its resonance frequency, its gain and losses, and its degree of coupling with the Source mode (whose intensity is referred as a$_S$). When mode S is not pumped, the coupling term is negligible and mode D oscillates at its fundamental frequency. On the other hand, if energy in the larger mode is enough, one may search for a solution oscillating at the frequency $\omega_S$ of the S mode. In this case after  simple algebra (see supplementary informations) one finds that
\begin{equation}
A=\frac{Energy\; of\;D}{Energy\; of\;S@D}=\frac{K^2}{\left[(\alpha_D-g_D)^2+(\omega_D-\omega_S)^2 \right]}.
\label{lorentian}
\end{equation}
Equation (\ref{lorentian}) shows that amplification follows a Lorentian lineshape when increasing the gain for the drain g$_D$ (see figure \ref{fig:figure2} d), and reach a maximum value $A=[K/(\omega_D-\omega_S)]^2$. Knowing maximum amplification and the frequency difference $\omega_D-\omega_S=\Delta \omega$, one can use the equation (\ref{lorentian}) to estimate the K constant. At saturation and defining  $K_\lambda=\frac{\lambda K}{2 \pi c}$ one obtains from equation (\ref{lorentian}):
\begin{equation}
A=\frac{K^2}{\Delta\omega^2}=\frac{K_\lambda^2}{\Delta\lambda^2}
\label{saturation}
\end{equation}
where $\Delta \lambda$ is the modes difference in wavelength and K$_\lambda$ is the coupling measured in the units of wavelenght. In the case reported in figure  \ref{fig:figure2} the maximum amplification has a value of $\sim7$ and $\Delta \lambda$ is 4.4 nm resulting in a coupling constant of $K_\lambda\sim$ 11 nm, which is a number close to the one obtained by fitting the experimental data with equation (\ref{lorentian}) (see details in supplementary material). $K_\lambda$ is thus the spectral distance for two modes that allows for a maximum amplification of 1; spectrally closer modes may achieve higher amplifications while more distant ones will be not amplified exactly like in the case of an mode forced to oscillate far from its resonance.

The range over which spatial transfer from S to D is possible can be estimated by measuring how closely a \emph{Drain} spectrum with \emph{Source} ON resembles its natural state (Source OFF) as a function of distance. Figure \ref{fig:figure4} shows the degree of interaction (defined in the SI) as a function of \emph{Gate} to \emph{Source} distance $\Delta X$ for cluster C1. The interaction strength drops to 0 at distances larger than 15 $\mu$m. The top panels in figure \ref{fig:figure4} show the intensity distribution retrieved with various $\Delta X$: for small $\Delta X$ resonances (hotspots in the images) of D and S dwell in the same region thus energy may flow freely between different modes. When $\Delta X$ is larger a region with low pumping (and thus large losses) separates the two sets of hotspots impeding the transport.

In summary we have found that lasing modes in a RL can be made to interact so that not only energy, but spectral features can be actively controlled and transported to remote locations by the activation of a second light signal much like electronic signals do. Above a world of possible future applications for controlling light flow, the formulation of a consistent model for the interaction of  modes, and the first measurement of the inter mode coupling constant is a milestone which will allow a more thorough understanding of its underlying physics and more complete comprehension of the nature of switching between different RL regimes.



\noindent {\it Acknowledgments ---}
The research leading to these results has received funding from the ERC under the EC's Seventh Framework Program (FP7/2007-2013) grant agreement n.201766, EU FP7 NoE Nanophotonics4Enery Grant No 248855; the Spanish MICINN CSD2007-0046 (Nanolight.es); MAT2009-07841 (GLUSFA) and Comunidad de Madrid S2009/MAT-1756 (PHAMA). \emph{Competing Interests} The authors declare that they have no competing financial interests.
\emph{Correspondence} Correspondence and requests for materials should be addressed to  marco.leonetti@roma1.infn.it
\emph{Contribution} Marco Leonetti designed and performed experiments, Cefe Lopez and Claudio Conti participated in data interpretation and in writing the paper.

\end{document}